# New insight in cervical cancer diagnosis using convolution neural network architecture

**Ach Khozaimi[1,2], Wayan Firdaus Mahmudy[3]**

[1]Department of Computer Science, Faculty of Engineering, Universitas Trunojoyo, Madura, Indonesia
[2]Department of Mathematics, Faculty of Mathematic and Natural Sciences, Brawijaya University, Malang, Indonesia
[3]Department of Informatics Engineering, Faculty of Computer Science, Brawijaya University, Malang, Indonesia

*Corresponding Author:*

Wayan Firdaus Mahmudy
Department of Informatics Engineering, Faculty of Computer Science, Brawijaya University
St. Veteran No.10-11, Ketawanggede, Lowokwaru, Malang 65145, Indonesia
Email: wayanfm@ub.ac.id

## 1. INTRODUCTION

Cancer is the second leading cause of death worldwide after heart attack [1]. The global cancer observatory (GCO) reported that in 2018, there were an estimated 570,000 new cases of cervical cancer and 311,000 deaths from the disease. Without efforts to reduce incidence rates, cervical cancer will kill an estimated 460,000 people annually by 2040 [2]. Approximately 90% of cervical cancer deaths occur in the developing world [3]. In Indonesia, cervical cancer is the second leading cause of death after breast cancer [4]. Cervical cancer screening is one of the efforts to reduce the risk of death [5]. Cervical cancer screening is aimed at detecting abnormal cell growth and proliferation. Malignant cells that invade one organ have the potential to spread to other organs [6]. Papanicolaou smear is an important screening method for detecting and classifying cervical cancer [3].

There are five steps in the process of machine-learning-based Pap smear analysis: image acquisition, preprocessing, segmentation, feature extraction, and classification [7]. At each stage, the following algorithms are required: histogram equalization [8] for preprocessing and Perona Malik diffusion for noise reduction [9]; global thresholding [10] for segmenting the region of interest; fuzzy c-mean for feature extraction; and k-nearest neighbour (KNN), neural network, and support vector machine (SVM) [11] for the classification process. The convolutional neural network (CNN) model shown in Figure 1 can be used to trim the five stages. The CNN is designed to process image data by identifying patterns in the image. The CNN architecture consists of convolutional and pooling layers. This is followed by several fully connected layers [12].

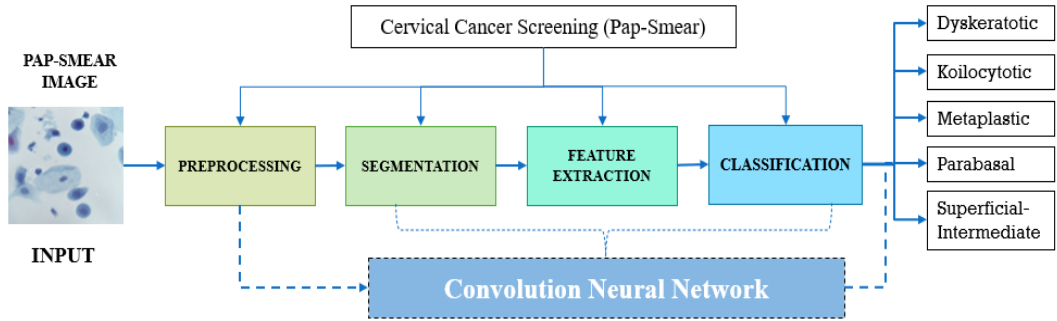

Figure 1. Cervical cancer classification with and without CNN

Some investigations have been carried out regarding the identification of cervical cancer, such as Chauhan and Singh [13]. Using the CNN model for cervical cancer detection, this study provides 96.89% accuracy, 93.38% precision, 93.75% sensitivity, and 94.15% F-score. The hybrid of CNN with classification algorithms has also been the subject of other studies. For example, Jia et al. [3] used a CNN-SVM model to detect cervical cancer. Research has been conducted on the combination of CNNs with artificial neural networks (ANNs) and KNNs [14]. In addition to a hybrid of CNN and classification methods, some researchers have also improved the performance of CNN by comparing optimizers on CNN. For example, a study conducted by [15] modified an existing CNN optimizer using the stochastic gradient descent (SGD), Adam, and root mean square propagation (RMSprop) optimizer models, in which the SGD outperformed the Adam and RMSProp optimizer models. Singh et al. [16] compared the performance of Adam, Adadelta, and SGD optimizer models on CNN for leukocyte classification, and the Adam optimizer emerged as the winner in the comparison. Mustafa and Dauda [17] compared the performance of optimizers on a CNN for cervical cancer classification. The optimization models compared were SGD RMSprop, and adaptive moment estimation (Adam). From the test results, it was concluded that the Adam optimizer had the best accuracy value compared to the other two optimizer models. Unlike previous studies that compared only three optimizer models, Dogo et al. [18] conducted a study comparing gradient-based optimizer models, namely, SGD, vSGD, SGDm, SGDm+n, RMSProp, Adam, AdaGrad, AdaDelta, Adamax, and Nadam. The experimental results show that Nadam performs better than the other optimizer models, whereas AdaDelta exhibits the worst performance.

Based on previous studies, the selection of a suitable optimizer depends on the specific case study and nature of the images being considered. In light of this understanding, the present study undertakes a systematic investigation and comparison of optimizer performance in the context of CNNs tasked with the classification of cervical cancer Pap smear medical images using the SipakMed dataset. The CNN architectures investigated include Resnet 18, Resnet 34, and VGG16, both with and without applying transfer learning (TL). The optimizer models include a comprehensive range of SGD, RMSProp, Adam, AdaGrad, AdaDelta, Adamax, and Nadam. This study aims to identify the most effective optimizer for improving the classification performance of CNNs in the domain of cervical cancer Pap smear image recognition. It provides valuable insights into the optimal model configuration for medical image analysis.

## 2. METHOD

This study used a dataset of Pap smear images derived from Kaggle, namely the SipakMed dataset. The SipakMed image is divided into five classes based on the degree of malignancy of cervical cancer and consists of 4049 cropped cell images. This dataset was then divided into three parts: 80% of the training data, 10% of the test data, and 10% of the validation data. Preprocessing was performed on the Pap smear images. It consists of image resizing to 224×224 pixels and image augmentation.

This study uses three CNN architectures, VGG16, Resnet 18, and Resnet 34, with and without TL. The RMSProp, Adam, SGD, Adadelta, Adagrad, Adamax, and Nadam optimizer techniques are compared in this study. The parameter settings for training the CNN model included a batch size of 16, random seed of 1, 30 epochs, and learning rate of 0.001. The performance metrics used in this research were the accuracy and loss function. The experimental method is illustrated in Figure 2. The software used is Python version 3.10 with a Jupyter notebook installed on a Windows 11 operating system. The hardware used was an AMD Ryzen 5 5500 CPU, 32 GB of RAM, 12 GB of Nvidia GeForce RTX 3060 graphics card, and 512 SSD.

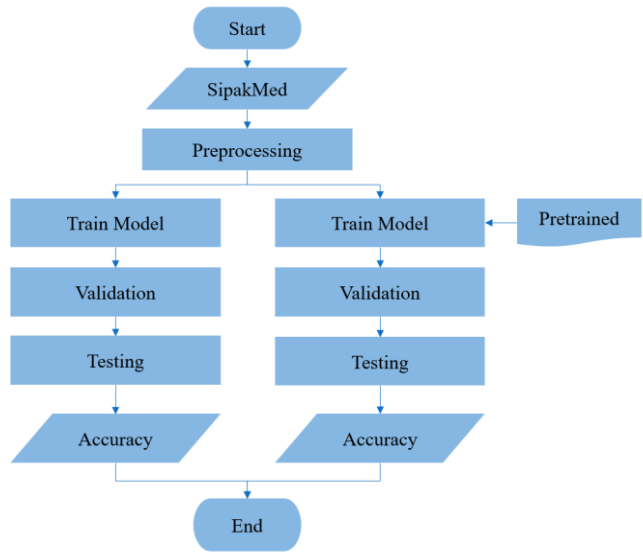

Figure 2. Experimental methods

## 2.1. Convolution neural network

CNN are a specific type of artificial neural network that can handle structured data, particularly images [14], [19]. CNNs share a structural resemblance with regular feedforward neural networks. The CNN architecture consists of a convolutional layer, pooling layer, and fully connected layers [12]. Figure 3 shows the arrangement of the CNN layers. However, they possess additional layers that enhance their proficiency in addressing computer vision problems. CNNs have been widely and successfully used to classify medical images [13], [20]. CNN modifications have also been performed to improve model performance, such as hybrid CNN models [21]–[25], CNN layer arrangement [25], [26], and CNN hyperparameter optimization [27]–[30].

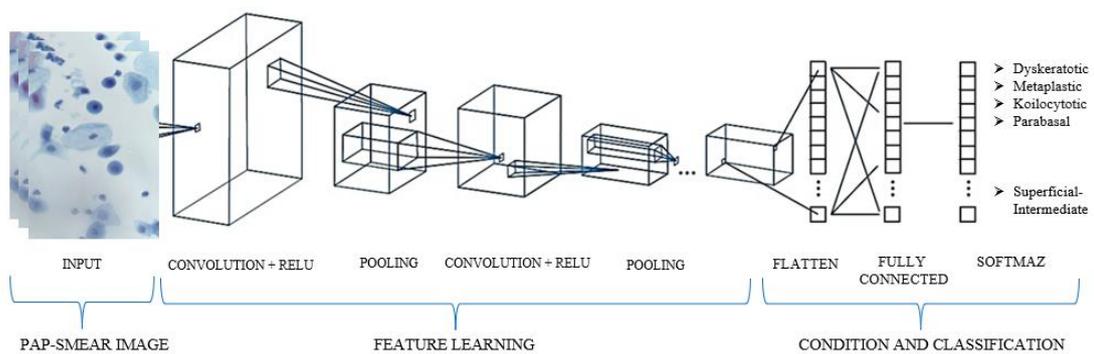

Figure 3. CNN architecture for cervical cancer classification

When implementing a CNN, optimization techniques are used to optimally manage network learning by minimizing the loss function of the CNN, thereby improving training efficiency [31]. This helps CNN networks learn quickly and efficiently and achieve better results in image recognition tasks. Some optimization techniques for implementing CNN learning include SGD, RMSprop, Adam, AdaGrad, AdaDelta, Adamax, and Nadam.

## 2.2. Stochastic gradient descent

The SGD is a fundamental optimization method. It has been employed in neural networks. SGD utilizes the gradient of slack function J with respect to each parameter to update the model. This was achieved by altering the weights of the parameters during each iteration [31]. SGD updates the parameters on each training data $x^{(i)}$ serta label $y^{(i)}$ with the following formula, where $\eta$ is the learning rate as (1).

$$\theta = \theta - \eta . \nabla \theta J(\theta; x^{(i)} y^{(i)}) \tag{1}$$

The SGD has several advantages. These include computational efficiency, fast convergence, the regularization effect, and the ability to escape from local minima. The SGD has several important applications. These methods include deep learning, natural language processing, and binary classification. The weakness of this technique is that it is sensitive to the learning rate, can be trapped in a local minimum, is inconsistent, susceptible to noise, and requires continuous updating of the weight values [16]. In addition to these weaknesses, SGD has several advantages. First, SGD is fast because it does not use the entire dataset for training, thereby increasing computational speed. Second, to minimize the occurrence of local minima, SGD can adaptively modify the estimation of the gradient matrix for each parameter, both in terms of first- and second-order, based on the loss function [32]. Several studies have used SGD for CNN optimizers [33]–[36].

## 2.3. Root mean square propagation

RMSProp is an optimization approach employed for training neural networks. This approach is an extension of gradient descent optimization that incorporates a decaying average of partial gradients to adjust the step size for each parameter. Let's assume you have a parameter $\theta$ that you are optimizing, and $g_t$ is the gradient of the loss function with respect to $\theta$ at time step $t$. The update rule for RMSprop is as (2) and (3).

$$E[g^2]_t = 0.9 E[g^2]_{t+1} + 0.1 g_t^2 \tag{2}$$

$$\theta_{t+1,i} = \theta_{t,i} - \frac{\eta}{\sqrt{\mathcal{E}_{t+1,i}^{g \circ g} + \varepsilon}} \cdot g_{t,i}, \ \forall t \tag{3}$$

$E[g^2]_t$ is the exponentially decaying average of past squared gradients., $\eta$ learning rate, $\epsilon$ is a small constant added for numerical stability (usually around) $10^{-8}$. The algorithm adapts the learning rate for each parameter based on the magnitude of recent gradients. If a parameter has been receiving large gradients, its effective learning rate is reduced, and vice versa. This aids in faster convergence and more stable training [36], [37].

## 2.4. Adaptive moment estimation

Adam optimizer is a widely used optimization technique in neural network training. The Adam algorithm integrates concepts from the RMSprop and momentum algorithms to offer a dynamic learning rate and momentum for every parameter. This enhances the efficacy of gradient-based optimization methods. Adam calculates and tracks two types of moving averages for each parameter: the first moment, which represents the mean, and the second moment, which represents the uncentered variance [38], [39]. The purpose of these moving averages was to dynamically modify the learning rates for each parameter. Formula for Adam:

Let $t$ be the current iteration, $\alpha$ is learning rate, $\beta_1$ and $\beta_2$ be the decay rates for the moment estimates, and $\epsilon$ be a small constant for numerical stability (usually set to $10^{-8}$). Initial first moment estimate ($m_0 = 0$); $v_0 = 0$ (initial second moment estimate), $t = 0$. The update rule for Adam is as (4) and (5):

$$m_t = \beta_1 . m_t - 1 + (1 - \beta_1). g_t \tag{4}$$

$$v_t = \beta_2 . v_{t-1} + (1 - \beta_2). g_t \tag{5}$$

$g_t$ is the gradient of the loss with respect to the parameter at time $t$.

$$m_t = m_t / 1 - \beta_1^t \tag{6}$$

$$v_t = m_t / 1 - \beta_2^t \tag{7}$$

This step corrects the bias introduced by the moving averages at the beginning of training. In (6) and (7) used for bias correction as (8).

$$\theta_{t+1,i} = \theta_{t,i} - \frac{\eta}{\sqrt{v_t + \varepsilon}} . m_t \tag{8}$$

$\theta$ is the parameter being optimized. In (8) is used to update parameter. Adam adjusts the learning rate for each parameter by considering both the average (mean) and spread (uncentered variance) of the gradients. The algorithm combines the advantages of RMSprop and momentum to provide efficient optimization for a wide range of deep-learning tasks [40], [41].

### 2.5. Adagrad

Adagrad, also known as the adaptive gradient, is a frequently employed optimization strategy in machine learning, specifically for training neural networks. Adagrad adjusts the learning rate of each parameter by proportionally scaling them according to the past squared gradients of the parameters. Consequently, parameters linked to rare features receive more substantial adjustments, whereas those linked to common features receive more modest updates [42]. Adagrad maintains a per-parameter learning rate that adapts over time based on the historical gradient information for each parameter. Formula for Adagrad as (9):

$$\theta_{t+1} = \theta_t - \frac{\eta}{\sqrt{G_t + \varepsilon}} \cdot g_t \quad (9)$$

$g_t$ is a diagonal matrix where each diagonal element corresponds to the sum of squared gradients for a particular parameter up to time $t$. $\eta$ is the learning rate.

Adagrad adjusted the learning rates by considering the past squared gradients of each parameter. This results in more substantial updates to parameters linked to rare features, and fewer updates to parameters linked to common features. However, one drawback of Adagrad is that the learning rates may become very small over time, leading to a slow convergence. This issue has led to the development of other adaptive optimization algorithms such as RMSprop and Adam [43].

### 2.6. AdaDelta

Adadelta is an optimization algorithm designed to address some of the limitations of Adagrad, specifically the issue of continually decreasing learning rates. Adadelta dynamically adjusts the learning rates during training, allowing it to adapt to the changing characteristics of the optimization landscape. Adadelta employed a method to calculate the average of the squared gradients from previous iterations to dynamically adjust the learning rates for each parameter [16], [44]. The algorithm eliminates the need for an initial learning rate and tracks the running average of the squared parameter updates.

$$RMS[g^2]_t = E[g^2]_t = \beta . E[g^2]_{t+1} + (1 - \beta) . g_t^2 \quad (10)$$

$E[g^2]_t$ is a running average of squared gradients.

$$\Delta \theta_t = - \frac{RMS|\Delta\theta|_{t+1}}{RMS|g|_t} \cdot g_t \quad (11)$$

This (11) is for compute update value.

$$\theta_{(t+1)} = \theta_t + \Delta \theta_t \quad (12)$$

This formula (12) is used to update parameter.

Adadelta dynamically adjusts the learning rates based on past squared gradients and the past squared parameter updates. This aids in alleviating the problems of rapidly increasing or decreasing learning rates and enables a more stable and efficient training of neural networks. Some studies used the Adadelta optimizer for the CNN model [34], [43].

### 2.7. Adamax

Adamax is a modified version of the Adam optimization method that specifically targets the L-infinity norm (maximum norm) of the gradients in the update rule. Adamax, like Adam, calculates the moving averages of the first moment (mean) and the exponentially decaying average of the infinity norm of the gradients. Adamax adjusts the learning rates for each parameter by considering both the mean and maximum norms of the gradients. Let $t$ be the current iteration, $\theta$ be the parameter being optimized, $g_t$ be the gradient of the loss with respect to $\theta$ at time $t$, $\beta_1$ and $\beta_2$ be the decay rates for the moment estimates.

$$\theta_{t+1} = \theta_t - \frac{\eta}{u_t} . m_t \quad (13)$$

$$u_t = \beta_2^\infty . v_t + (1 - \beta_2^\infty) . |g_t|^\infty = max(\beta_2 . v_{t-1}, |g_t|) \quad (14)$$

η is learning rate.

The advantage of Adamax over Adam, especially when dealing with sparse gradients or very noisy data, is that it is more stable and robust. In addition, Adamax requires less memory because it does not store the squared gradients. This makes it more memory efficient [45]. In CNNs, Adamax can help speed up convergence during training by adjusting the learning rates according to the size of the gradients, which is particularly beneficial when dealing with complex datasets with varying levels of feature importance [46].

### 2.8. Nadam

Nadam is an optimization algorithm that merges concepts from the nesterov accelerated gradient (NAG) and Adam optimizer. This is an abbreviation for the nesterov-accelerated Adam estimation. This incorporates the advantages of the nesterov momentum and adaptive learning rates. Nadam is a variant of the Adam optimization algorithm that integrates Nesterov momentum into the parameter update process. This modification aims to improve the convergence speed and generalization. This is the formula for the Nadam optimizer.

$$\theta_{t+1} = \theta_t - \frac{\eta}{\sqrt{G_t}+\varepsilon} \cdot (\beta_1 \cdot m_t + \frac{(1-\beta_1) \cdot g_t}{1-\beta_1^t}) \tag{15}$$

$\theta$ be the parameter being optimized, $g_t$ be the gradient of the loss with respect to $\theta$ at time $t$, $\beta_1$ and $\beta_2$ be the decay rates for the moment estimates, $\epsilon$ be a small constant for numerical stability (usually set to $10^{-8}$), and η be the learning rate. Nadam incorporated the Nesterov momentum into the update rule of the Adam optimizer. This combination is intended to provide faster convergence and better handling of saddle points compared with standard momentum methods. This is particularly useful for training deep neural networks [47].

## 3. RESULTS AND DISCUSSION

In this study, we investigated the performance of different optimizers in cervical cancer diagnosis using CNNs and compared their efficacy. Cervical cancer remains a significant global public health concern, and its early detection plays a crucial role in improving patient outcomes. CNNs have shown promise in automating the diagnosis process, but the choice of optimizer can significantly affect model performance. Optimizer performance test results for cervical cancer detection using CNN on VGG-16, ResNet-18, and ResNet-34 architectures with and without TL are as follows:

### 3.1. General result

In addition, we were able to determine which of these models had been trained on more data than those used in our study (SipakMed), so we were able to determine which of these models had been trained on more data than those used in our study (SipakMed). The results of this test also show that the test for the TL CNN architecture is stable for all optimizer types used. In other words, if we use a trained model or TL, our optimizer will not have a significant effect on that model.

### 3.2. VGG-16

The test results for the VGG-16 architecture are listed in Table 1. The Nadam optimizer with TL achieved an accuracy value of 88.6%. This was followed by Adamax, with an accuracy of 85.6%. When testing the VGG16 model without TL, the Adamax optimizer performed better with an accuracy value of 66.8%, whereas the second position was occupied by Nadam with an accuracy value of 66.1%. The difference in the accuracy between Nadam and Adamax was 0.7%. The test results on this VGG16 architecture show that adadelta performs worst for models without TL, with an accuracy value of 0.205, whereas for models using TL, RMSProp performs worst, with an accuracy value of 0.730.

Table 1. Result simulation for VGG-16 architecture

| Metric | RMSProp | Adam | SGD | Adadelta | Adagrad | Adamax | Nadam |
|---|---|---|---|---|---|---|---|
| Accuracy | 0.594 | 0.656 | 0.552 | 0.205 | 0.653 | 0.668 | 0.661 |
| Lose | 0.034 | 0.036 | 0.040 | 0.050 | 0.040 | 0.034 | 0.032 |
| Accuracy (TL) | 0.730 | 0.854 | 0.809 | 0.772 | 0.849 | 0.856 | 0.886 |
| Lose (TL) | 0.086 | 0.027 | 0.018 | 0.028 | 0.012 | 0.014 | 0.016 |

### 3.3. Resnet-18

The next test was performed on the Resnet-18 architecture, the results are presented in Table 2. Adamax, with accuracy values of 0.884 and 0.728, respectively, had the best performance compared with the other optimizers in both the Resnet-18 model with and without TL. Adam was then used for the model with TL with a value of 0.879, and Adagrad was used for the model without TL with a value of 0.683.

Table 2. Result simulation for Resnet-18 architecture

| Metric | RMSProp | Adam | SGD | Adadelta | Adagrad | Adamax | Nadam |
|---|---|---|---|---|---|---|---|
| Accuracy | 0.205 | 0.619 | 0.641 | 0.644 | 0.683 | 0.728 | 0.426 |
| Lose | 0.051 | 0.034 | 0.027 | 0.045 | 0.027 | 0.026 | 0.043 |
| Accuracy (TL) | 0.676 | 0.879 | 0.871 | 0.708 | 0.879 | 0.884 | 0.748 |
| Lose (TL) | 0.024 | 0.014 | 0.136 | 0.037 | 0.017 | 0.014 | 0.022 |

### 3.4. Resnet-34

Table 3 presents a detailed comparison of various optimization algorithms based on the key performance metrics for the Resnet-34 architecture. The metrics include accuracy, which reflects the proportion of correctly classified instances and losses and serves as a measure of the model's error. Among the featured optimizers, Adam had the lowest accuracy at 0.205 and a corresponding loss of 0.051. By contrast, Adagrad boasts the highest accuracy at 0.507, coupled with a relatively low loss of 0.040. Notably, the TL results indicate how well these optimizers adapt to a related task. Adam scenarios with an accuracy (TL) of 0.850 and a minimal loss of 0.014, reinforcing its robustness across diverse tasks. Conversely, Nadam exhibited a significant increase in loss during TL, thereby highlighting its potential limitations. These findings underscore the critical role of optimizer selection in shaping model performance, with Adam emerging as a favorable choice for both the initial task and TL contexts. Although slightly less accurate than Nadam, Adamax is an optimizer that can be an option for the Resnet-34 architecture without TL.

Table 3. Result simulation for Resnet-34 architecture

| Metric | RMSProp | Adam | SGD | Adadelta | Adagrad | Adamax | Nadam |
|---|---|---|---|---|---|---|---|
| Accuracy | 0.311 | 0.205 | 0.234 | 0.453 | 0.507 | 0.540 | 0.574 |
| Lose | 0.048 | 0.051 | 0.050 | 0.044 | 0.040 | 0.039 | 0.034 |
| Accuracy (TL) | 0.757 | 0.850 | 0.860 | 0.710 | 0.824 | 0.820 | 0.821 |
| Lose (TL) | 0.025 | 0.014 | 0.014 | 0.037 | 0.015 | 0.015 | 0.160 |

## 4. CONCLUSION

This study investigated the importance of optimizer selection in CNN cervical cancer image classification models. The results highlight the importance of TL, showing consistently better performance across all CNN topologies and optimization techniques. Adamax was shown to be the most successful optimizer, achieving the highest accuracy of 72.8% on the VGG-16 architecture and 66.8% on the Resnet-18 design. On Resnet-34, Adamax achieved an accuracy of 54.0% but was slightly outperformed by Nadam by 0.034%. Adamax is a reliable and effective optimizer for CNNs in the classification of cervical cancer using Pap smear images on Resnet-18, Resnet-34, and VGG-16 architectures. Future research can investigate the effect of noise and contrast on the performance of optimizers by reducing the noise and increasing the contrast in the Pap smear images.

**ACKNOWLEDGEMENTS**

Ach Khozaimi, BPI ID 202327091034, would like to thanks the Ministry of Education, Culture, Research, and Technology of the Republic of Indonesia through the Center for Higher Education Fund (BPPT) and Indonesia Endowment Funds for Education (LPDP) for providing the Indonesian Education Scholarship (BPI) under No. 00044/BPPT/BPI.06/9/2023.

## BIOGRAPHIES OF AUTHORS

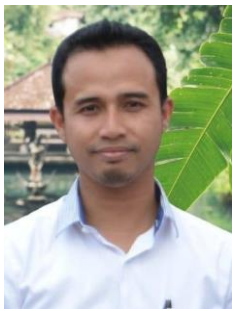
**Ach Khozaimi** is a lecturer in informatics engineering at Trunojoyo University of Madura. Obtained bachelor degree in informatic from Trunojoyo University of Madura, Indonesia, master degree in informatic from Institut Teknologi Sepuluh November (ITS), Indonesia. Today is currently pursuing his doctorate at Brawijaya University, Indonesia, in the Department of Mathematics. The doctoral programmed is funded by the Higher Education Financing Centre (BPPT) of the Ministry of Education and Culture Research Technology of the Republic of Indonesia. He can be contacted at email email: khozaimi@trunojoyo.ac.id.

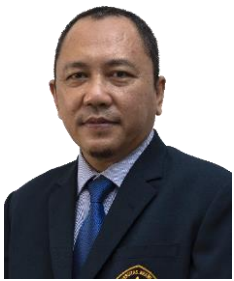
**Wayan Firdaus Mahmudy** obtained bachelor degree in mathematics from University of Brawijaya, Indonesia, master degree in information technology from Institut Teknologi Sepuluh November (ITS), Indonesia, and completed his Ph.D. in Manufacturing Engineering at University of South Australia. He is a lecturer at Department of Computer Science, University of Brawijaya (UB), Indonesia. His research interests include optimization of combinatorial problems and machine learning. He can be contacted at email: wayanfm@ub.ac.id.